\documentclass[sigconf,nonacm]{acmart}
\AtBeginDocument{%
  }

\setcopyright{none}
\usepackage{xcolor}

\newif\ifdraft
\draftfalse

\ifdraft
  \newcommand{\todo}[1]{\textcolor{teal}{#1}}
  \newcommand{\change}[1]{\textcolor{brown}{#1}}
\else
  \newcommand{\change}[1]{#1}
  \newcommand{\todo}[1]{#1}
\fi

\begin{document}

\title{Exploring 3D Glyph Physicalizations for Public Engagement through River Health}

\author{Maria Teresa Ortoleva}
\email{maria-teresa.ortoleva@kcl.ac.uk}
\orcid{0009-0004-0610-5425}
\affiliation{%
  \institution{King's College London}
  \city{London}
  \country{United Kingdom}
}

\author{Min Chen}
\email{min.chen@oerc.ox.ac.uk}
\orcid{0000-0001-5320-5729}
\affiliation{%
  \institution{University of Oxford}
  \city{Oxford}
  \country{United Kingdom}
}

\author{Rita Borgo}
\email{rita.borgo@kcl.ac.uk}
\orcid{0000-0003-2875-6793}
\affiliation{%
  \institution{King's College London}
  \city{London}
  \country{United Kingdom}
}

\author{Alfie Abdul-Rahman}
\email{alfie.abdulrahman@kcl.ac.uk}
\orcid{0000-0002-6257-876X}
\affiliation{%
  \institution{King's College London}
  \city{London}
  \country{United Kingdom}
}

\renewcommand{\shortauthors}{Ortoleva et al.}

\begin{abstract}
\textbf{Introduction.} 
In this paper, we present the preliminary design of a toolkit for making glyph-based physicalizations for public engagement. We use London river health data as a case study: a data set of significance to urban issues related to climate change and of interest to draw public attention, as part of the Greater London Authority's strategies.
\textbf{Design.} We present the components of a 3D glyph-making toolkit, its encodings, and a step-by-step process for crafting a physicalization of a river's water quality using recycled materials. We reason about how users can use the template to learn about a data set while reflecting on the data's significance to their personal experience and self-mapping onto the physicalization. 
\textbf{Reflection.} We reflect on the opportunities that extending the design space of glyphs to 3D physicalization offers for supporting public engagement with complex, multi-dimensional data sets, scaffolding cognitive processes, and self-reflection, thereby bringing crucial environmental data to life. 
\textbf{Conclusion.} Future implementation of the 3D glyph template will enable the public of all abilities to explore river health data, physicalize complexity, and realize its relevance. We hope that its use in public engagement workshops will help raise awareness, invite care, and foster a sense of belonging.
\end{abstract}

%\begin{CCSXML}
%<ccs2012>
%<concept>
%<concept_id>10003120.10003145.10003146</concept_id>
%<concept_desc>Human-centered computing~Visualization techniques</concept_desc>
%<concept_significance>500</concept_significance>
%</concept>
%</ccs2012>
%\end{CCSXML}
 
%\ccsdesc[500]{Human-centered computing~Visualization techniques}

%\ccsdesc[500]{Human-centered computing}

%Must be 3 keywords.
%Publice Engagement in place of Glyphs?

\begin{teaserfigure}
  \includegraphics[width=\textwidth]{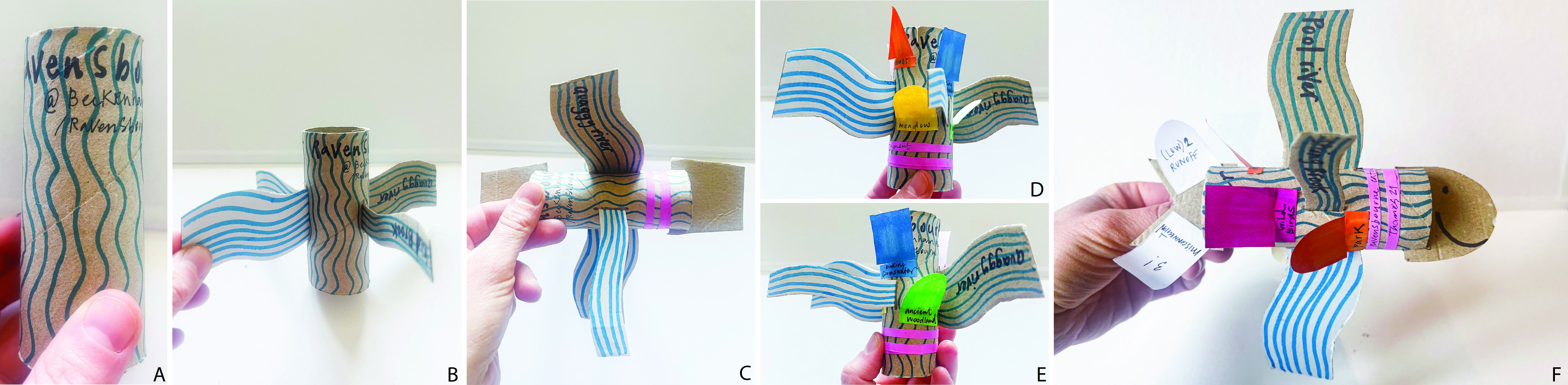}
  \caption{Step by step process of assembly of a 3D glyph physicalization mapping the catchment, river quality, and personal relationship of a participant towards the Ravensbourne River, in the Thames basin, London.}
  \Description{}
  \label{fig:teaser}
\end{teaserfigure}

\received{30 April 2026}
\received[accepted]{30 June 2026}

\maketitle

\begin{center}
\textit{This paper has been accepted for publication in the Visualising Climate 2026 Proceedings. 4-6 November 2026, Bologna, Italy.}
\end{center}

\section{Introduction}
Climate change is at \textit{``code red''} emergency level~\cite{Ripple:2022:warning-of-climate-emergency} with effects that impact a wide variety of environmental, ecological, and societal issues, as scientists warned at a global level~\cite{alliance-world-scientists-warnings}.
%2002 study on impact of climate change on London (GLA, Wilby etc)
While an increasing proportion of the world's population lives in cities, urban environments are strongly affected by climate change. 
Therefore, involving urban communities and promoting understanding of climate change are key aspects of adaptation and care strategies. 
Promoting direct, hands-on public engagement with data can offer critical help to raising necessary and informed awareness of environmental issues~\cite{ellwein_using_2014}. \todo{To this end, various works in data visualization research have explored opportunities in using physicalization strategies to} \change{spark reckoning about climate issues through the physical, often spectacular, manifestation of related data (e.g.~\cite{sauve_econundrum_2020, sauve_edo_2023, perovich_chemicals_2021, de_kreij_data_2024, aragon_risingemotions_2021}.} %promote a concrete understanding of environmental data (e.g.~\cite{sauve_econundrum_2020, sauve_edo_2023, perovich_chemicals_2021, de_kreij_data_2024, aragon_risingemotions_2021}).
\change{%MAKE MORE EXPLICIT WHY 3D GLYPHS ARE PARTICULARLY APPROPRIATE (critique A) 
In this paper, we contribute to these efforts a novel 3D glyph-based approach that builds on the aptitude of glyphs to represent complexity and embeds it into a participatory data physicalization protocol.
%, hence particularly apt to multivariate and interdependent environmental data.
%complex multivariate and interdependent data, such as environmental data intrinsically is. 
Hence, we present a 3D glyph physicalization making toolkit to promote public engagement and support lay people explore and %drilling into
concretely understand challenging multivariate and interdependent environmental data, thereby fostering  informed environmental awareness and inspiring changes in attitudes.}
%In this paper, we present a 3D glyph physicalization for a toolkit to support public engagement with complex environmental data, raise awareness of the issues they represent, and inspire changes in attitudes.

Our work takes the health of the London River as a case study. As a major world city spread across the catchment of the river Thames, London offers a relevant case study to observe the effects of climate change (pollution levels, river flow, reduced biodiversity, etc.~\cite{Jin:2012:hydrology-model-thames}) on urban water bodies and intervention strategies to improve both the cleanliness of the water and the welfare of its population% – one from which other cities globally can learn in their effort to adapt to climate change
~\cite{Clarke-GLA:2002:Londond-warming-report}.
Improving river health is an important focus of the 10-year Greater London Authority (GLA) intervention plan, which includes reducing water pollution, improving and expanding access to blue spaces, creating opportunities for citizens to engage with their water heritage, and preserving and increasing biodiversity~\cite{GLA:river-health-website}.
%\textit{``The Thames and its tributaries are the lifeblood of London, shaping communities, sustaining livelihoods and bringing people closer to nature.  [\dots] However, for too long they have been neglected and damaged by pollution. Five times more sewage flowed into London's rivers in 2023, compared to the previous year. Many species are in decline, despite efforts over the decades to restore them.''}~\cite{GLA:river-health-website}.
Engaging communities is key to the renovation of the London waterway, both as agents in maintaining clean water and as beneficiaries of improved water quality. 
The GLA's data-centered attitude provides informative data sets available to the public through the London Datastore~\cite{GLA:London-Datastore}; however, opportunities for engagement are needed to make the data sets known to audiences and bring them to life~\cite{web:londondatastore10}, support exploration of river heritage through data, raise awareness, and develop a sense of responsibility for their care and belonging. 

\todo{However, river quality data are a large, complex data set to understand and visualize, which presents challenges for both engaging a non-technical public and applying physicalization techniques.} \change{For this reason, in this work, we leverage the strength of glyph representations and,} taking London river health data as a case study, we explore %this paper explores 
the opportunities for public engagement offered by expanding glyph-based data representations into the 3D space of physicalization. \todo{In the 2D space, glyphs would offer a form of data visualization particularly apt at representing the multiple variables of a data set, and one that, by relying on figurative representation, symbols, and metaphors to create semantically resonant data representations, would provide easily recognizable representations for lay people. However, no correspondence of glyphs has been the direct object of exploration in the physicalization space and in hands-on approaches for a non-technical target group.}

%\textbf{\textit{Can we design a physical toolkit to engage a non-technical public in making 3D glyph-based data representations that explore London's river quality data? Can we support users in tracing personal connections to the data set and enrich their identity and sense of community around their local rivers?}}

In this short paper, we present the preliminary design of a 3D glyph-based data physicalization toolkit to help data visualization designers and facilitators create public engagement interventions that allow exploration and gain a concrete understanding of complex data about changes of climate, environment, ecology, and society %while fostering first-hand appreciation and understanding of data visualization
-- in this case, the Londoners' river heritage and water quality. The toolkit aims at supporting participants 
%take part and appreciate data representation technicques such as visual mapping, exploration of data overview and detail, perception and material encodings. Furthermore it supports reflecting on the significance of data, encouraging to incorporate own relationship, personal experience, and emotional response to blue spaces into the data set.
%Firstly, we experiment with the 3D-making of glyphs as a form of physicalization for public engagement. Secondly, we invite Londoners to explore their rivers' heritage and their water quality in the wake of climate change. We support them in 
working hands-on, exploratively, and playfully with the data, physicalizing them and integrating their own relationship, personal experience, and emotional response into the data set. In this paper, we present the grammar of a proposed glyph physicalization-making template and how its structure maps to our chosen data set: London river health data. \todo{%PROVIDE A MORE FOCUSED DEFINITION OF THE AUDIENCE (critique B)
} 
\change{While our design, at this preliminary stage, considers a generalized audience and situation, %still its framework remain adaptable
we further argue for opportunities to maintain its fundamental structure, while tailoring the visual and sensory aspects, materials, and degree of personalization to the characteristics of specific groups and learning contexts.
For example, future designers and facilitators may tune the toolkit to best engage a group of school children in a classroom, or vary it for local members of the community dropping in at a pop-up stand at a street festival, or adapt it for extended use with youth as part of a day-long workshop.}
Subsequent to this work, we aim to trial the toolkit with the attendees of the London Data Week 2026, a Greater London Authority public festival, that promotes data for everyone in London through free drop-in workshops, lectures, and displays hosted at participating venues across the city center and advertised widely to all ages, cultures, communities, and levels of ability and knowledge.

\section{Related Work}
\subsection{Glyph Data Visualizations}
Glyphs are \textit{``small independent visual objects that depict the characteristics of a data record''}, they are placed in space and use the language of visual signs, illustrative and dictionary-based schemes of encodings~\cite{Borgo:Glpyphs:2012}. Glyph design combines multiple visual channels (such as shape, color, texture, size, orientation, aspect ratio, curvature, \ldots) and draws on theories of semiotics, perception, and cognition to communicate data effectively. Thus, the strength of glyph-based visualization lies in its ability to represent complex, multi-dimensional data through readily perceived visual forms, which can be abstract and geometric, as well as illustrative, using symbols, metaphors, and figures that resonate semantically with the data they represent~\cite{van_koningsbruggen_metaphors_2024}. Thus, glyphs make for a vivid, approachable, and effective way of communicating complex data to lay users.

Glyphs are more commonly used and studied as elements for graphic visualizations. Yet, there are contexts in which glyphs are encountered as 3D objects. For example, maritime buoys, windsocks, or flags communicate through a dictionary-based code of shapes, colors, and geometric patterns (e.g., stripes) to an audience trained to recognize and read them. %Perhaps some physicalizations produced in data physicalization empirical studies could be re-looked at as glyphs??
However, the glyph, as a three-dimensional encoded structure, has not yet been directly explored in data physicalization research. 

%...yet data visualization research focuses on them as computer graphics elements of interest rather than as objects. There is therefore a research gap in the physicalization of glyphs which our work aims to fill. --- More on why is it relevant to explore 3D glyphs?

\subsection{Data Physicalization and Public Engagement}
%
%Data physicalization is documented to support the public in data literacy by exploring a data set and gaining knowledge of it~\cite{Jansen:2015:CHI}, reflecting on its value~\cite{thudt_data_2017}, following personal emotional connections with it~\cite{Wang:2019:Emotional-response}, and persuading its greater environmental and societal weight~\cite{Peng:2026:CHI:phys4good, aragon_risingemotions_2021}.
\textit{``A data physicalization is a physical artifact whose geometry or material properties encode data.''}~\cite{Jansen:2015:CHI} Physicalization can vary from everyday objects to art and design crafts to didactic and museum displays~\cite{wiki:dataphys}. They offer intermodal and perceptually attractive data experiences that are accessible to the public at all levels of data literacy, provide opportunities for embodied cognition~\cite{Jansen:2015:CHI}, reflectiveness~\cite{thudt_data_2017, karyda_narrative_2021}, and foster an emotional connection to data~\cite{Wang:2019:Emotional-response}. One benefit of creating and using material data representations is the opportunity they offer to engage the public in the hands-on crafting of their own data physicalizations, either in workshops~\cite{Huron:Lets-get-physical:2017} or through DIY toolkits that they can use independently and in their own space~\cite{Huron:2014:Constructive-vis}. In doing so, they make the data more ``graspable''~\cite{Huron:Lets-get-physical:2017}.

Data physicalization workshops are particularly useful for pedagogical purposes~\cite {hogan_pedagogy_2017} in exploring a data set and the principles of its representation. They help to build theoretical concepts and complex problems % -- such as air quality data --
more concretely explainable, for example, to children and youth~\cite{de_kreij_data_2024}. Collectively crafting, sharing, and reflecting on data helps build community, supports social interaction and reflection, and invests in meaning~\cite{Nissen:2015:data-things}. In private and domestic spaces, physicalization toolkits support independent, reflective work, most often with personal data~\cite{thudt_data_2017}. In the public realm, they create opportunities for public participation and motivate the social good~\cite{Peng:2026:CHI:phys4good}.

\change{Data physicalizations that communicate environmental data especially leverage the communicative, explanatory, and persuasive strength of physical data displays. Examples include} De Kreij et al.'s~\cite{de_kreij_data_2024} physicalization explaining air quality data to children. Sauv\'e et al.'s Edo~\cite{sauve_edo_2023} prompts reflection on the dietary impact on climate change through a large participatory display of people's food choices in a canteen, thereby visually translating the impact inherent in implicit choices. Similarly, Perovich et al.'s~\cite{perovich_chemicals_2021} work uses artistic spectacularization to reveal the invisible presence of pollution. Arag\'{o}n et al.'s~\cite{aragon_risingemotions_2021} examine the joint use of art and technology to persuasively display public emotions around environmental data and highlight the role of using emotions to help people realize significance. 

Our work in this paper aims to combine the strengths of hands-on
\change{creative engagement} with physical data visualization, \change{and with the representational and semantic opportunity of glyphs to digest complexity, thus} integrating both cognitive and emotional benefits. Firstly, our goal is to support users in exploring an external data set, tracking their learning, digging deeper, and thinking critically. Secondly, we aim to help them grasp the personal relevance of the data sets, incorporating self-mapping and approaches typical of personal data physicalization to create an emotional connection and a sense of belonging to the data topic. Hence, our goal is to use glyph physicalization as a place for combined cognitive and personal exploration.

%Matteo Moretti's PDP (Participatory Data Phys)~\cite{moretti_participatory_2020}

\section{The Data Set}
Through consultation with data experts at the GLA, we selected a significant data set worth supporting public attention towards, and interesting to explore through physical visualization. Our attention was drawn to the River Health Map (\url{https://apps.london.gov.uk/river-health/}), which allows exploration of river quality throughout the Thames basin in Greater London and provides links to a range of external resources to deepen understanding of the factors influencing water quality. We first focused on the descriptive data: composition of a river catchment (parent, river, tributary streams and other water basins), characteristics of the surrounding land (categories: plants, animals, use of land, natural spots of interest, reserves), partnerships and organizations looking after a river, points of public access, and recent renovations. We then focused on variables that affect rivers' risk of pollution: type of sewer (combined/separate), number of main roads and water runoff that ends up in the river, discharge, misconnections of sewage, and improvement works.

The GLA and London Data Store~\cite{GLA:London-Datastore} make London's data publicly accessible for consultation by stakeholders, as well as to the wider public. However, the lack of public awareness and the perception of data as too technical and unapproachable create the need for opportunities to increase public engagement~\cite{web:londondatastore10, web:londondatastore2025}. Although the map successfully visualizes complex technical information, it may remain mute to Londoners if it is disconnected from their direct experience of London rivers and from an understanding of how the facts and events recorded in these data affect them and their local area in the day-to-day. Creative and hands-on approaches to working with data, such as data-crafting workshops, can provide valuable tools for responding to these challenges. Building on this potential, we approach the River Health map through participatory 3D glyph physicalizations, with the aim of observing how they support efforts to get to know the city's rivers, improve their quality, and boost belonging to them.

\section{Designing a 3D Glyph Toolkit}

To design a data physicalization toolkit for making 3D glyphs that engage the public in exploring and reflecting on river health data, we worked on these fundamental principles: 
(i) the semantic resonance with the data set and the wider topic; 
(ii) the materials to use for the physicalizations; 
(iii) the structure and way of assembling/constructing; 
(iv) the mapping of the data characteristics to the physicalization features; and 
(v) the learning and reflecting process facilitated.

\begin{figure*}[ht]
    \centering
    \includegraphics[width=1\linewidth]{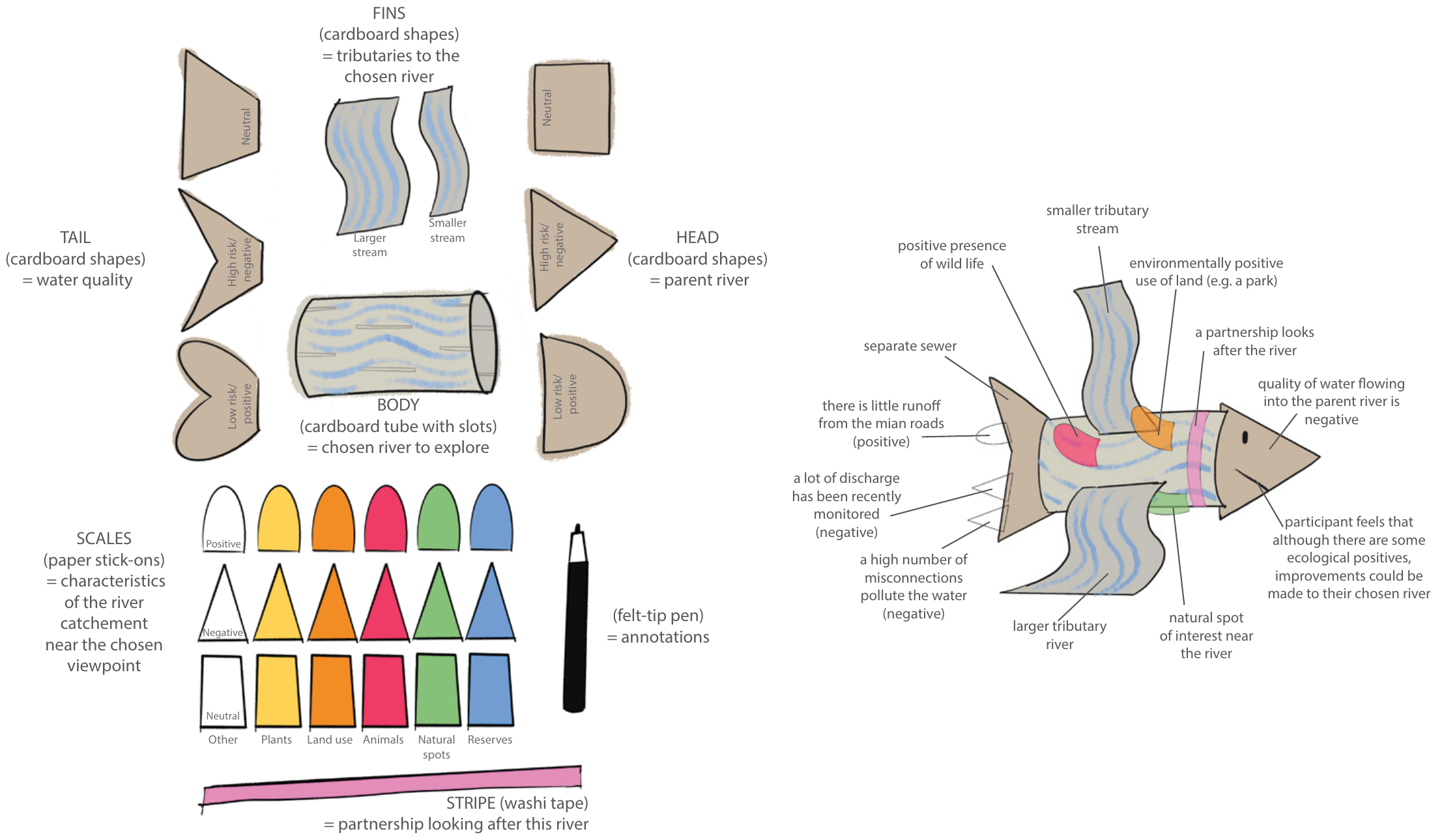}
    \caption{Left: The components of a 3D glyph physicalization toolkit: symbolic interpretation of each part, the materials, and the data they represent. Right: An example of a 3d glyph physicalization with the encodings explained.}
    \label{fig:toolkit-overview}
    \vspace{-10pt}
\end{figure*}

\paragraph{3D Glyph Shape and Slot-in Grammar}
Our design builds 3D glyphs around a central core to which additional geometric shapes are attached, as shown in Fig.~\ref{fig:toolkit-overview}. The mechanism follows that of slot-in toys, cut into 3D sheets of material and then assembled into a 3D structure. Animals and figurines are frequent themes in these children's play sets. Our chosen structure, while constructed from encoded geometric components, recalls the shape of a fish: a streamlined body (core), head and tail (attachments at the top and bottom), fins (lateral attachments around the core), and scales (stick-ons to the body). This is also a typical shape for underwater vehicles and marine objects that take advantage of the anatomy of fish for navigation. As such, the glyphs' shape and structure resonate semantically with the data set's water theme.
\begin{itemize}
    \item In our prototype, the body is a 10cm long, 4.5cm diameter paper tube, with pre-cut slots along its length and distributed around its circumference. Further cuts are made at the top and bottom.
    \item The next components of the kit are flat, 1.5-mm-thick corrugated cardboard shapes that slot into the body. They are divided into ``fins'' which slot into the side cuts, ``heads'' which slot into the top cuts, and ``tails'' which slot into the bottom cuts. 
    Heads and tails can have a square, rounded, or pointed ending.
    The shape's category, its size, and the type of ending encode information from the data set.
    \item Further paper flaps and tape strips are available to be stuck onto the body and the tail, and to map secondary data specifications. These scale-like attachments are color-coded to match different dimensions of the data set and come in squared, rounded, and pointed ends, like the heads and tails.
    \item Felt-tip pens allow for writing annotations directly onto the physicalization.
\end{itemize}

\paragraph{Materials}
Materials for the prototype are recycled cardboard in sheets and tubes, colored paper, tape, and felt-tip pens. Given the topic of the data set, we strived to select sustainable materials in tune with the environmental data to be represented. Furthermore, we envision the toolkit being distributed as instructions to different groups (from individuals to schools to community groups) for independent use. We therefore selected materials that are inexpensive and easy to source and craft, while also, of their own accord, serving as a commentary on environmental sustainability and encouraging the creative repurposing of domestic waste. For example, empty toilet rolls can serve as bodies, and flat cardboard heads, tails, and fins can be cut from delivery boxes or cereal boxes.

\paragraph{Mapping the 3D Glyph Grammar to the Characteristics of the River Health Data Set}
From the River Health data set, we created a list of steps to guide users in exploring the variables featured on the map and the external websites it links to. The prompts scaffold their learning, help them reflect on the significance of the data, and trace personal connections to it as they construct their physicalization. In Table~\ref{tab:encodings}, we present the steps for examining the data and physicalizing each characteristic as a glyph. In the process, we describe the encodings of our design and their correspondence to the data variables. We illustrate each step, providing an example of the physicalization of the Ravensbourne River's glyph that we created as a test (see Fig.~\ref{fig:teaser}).

%\begin{figure}[t]
%    \centering
%    \includegraphics[width=1\linewidth]{trial_ravensbourne_step-by-step_placeholder.pdf}
%    \caption{Step by step process of assembly of a trial 3D glyph -- figure is a placeholder and needs improving}
%    \label{fig:step-by-step}
%\end{figure}

\begin{table*}[ht]
    \centering
    \scalebox{0.9}{
    \begin{tabular}{|p{4cm}|p{3cm}|p{4cm}|p{3cm}|}
    \hline
       \textbf{Prompt}  & \textbf{Reflection} & \textbf{Physicalization} & \textbf{Example} \\\hline
    \hline
        Choose a river to explore and locate it on the map \textit{(e.g., your local river)} 
        & \textit{What is your relationship to this river?} 
        & Pick a core paper tube and label it with the name of your chosen river (Fig.~\ref{fig:teaser}A)
        & \textit{``Ravensbourne River''} (Fig.~\ref{fig:teaser}A) \\
    \hline
        Choose your viewpoint
        & 
        & Annotate on the core
        & \textit{``Near the entrance to Beckenham Place Park''} \\\hline
    \hline
        \textbf{Get to know your river: explore its catchment}
        &
        & \textbf{Work on the body}
        & \\
    \hline
        How many tributaries merge into your chosen river?
        Can you name them?
        &
        & Pick one fin shape per affluent river (larger/thinner fins to bigger/smaller streams). Label each fin with the tributary name and slot them into the tube.
        & 4 tributaries: Pool River (large), Quaggy River (large), Kid Brook (thin), and Spring Brook (thin). (Fig.~\ref{fig:teaser}B) \\ 
    \hline
        What is the main stem of your river?
        &
        & Pick a square head. Annotate the main stem name and insert it at the top.
        & \textit{``Thames (at Deptford Creek)''} \\
    \hline
        What are the characteristics of the land surrounding your river at your chosen viewpoint? \textit{(Explore through external links via the River Health map)}
        &
        & Use color-coded stick-ons to mark land use, animals, plants, nature reserves, spots of naturalistic interest, and others.
        Choose the appropriate ending for neutral, negative, and positive features. See Fig.~\ref{fig:toolkit-overview}. Annotate and stick onto the body.
        & Orange square: residential houses;
        orange dot: park;
        orange triangle: industrial area;
        red dot: wild birds;
        green dot: groundwater reserve;
        blue dot: ancient woodland.
        (Fig.~\ref{fig:teaser}D-E)\\
    \hline
        Who looks after your river? \textit{(What partnership?)}
        & \textit{Have you ever taken part in activities to look after your river?}
        & Add rings of washi tape and annotate the partnership name.
        & 1st ring: Thames21;
        2nd ring: Ravensbourne Catchment Improvement Group.
        (Fig.~\ref{fig:teaser}C)\\
    \hline
        What are the points of access to the community?
        & \textit{How do you access your river?
        What activities do you do?}
        & Add further color-coded stick-ons.
        & White stick-on: footpath along the river. Note: \textit{``Long walks''}\\\hline
    \hline
       \textbf{How healthy is your river?}
       & 
       & \textbf{Work on the tail}
       & \\
    \hline
       What sewer type is in the area?
       &
       & Combined sewer: insert a rounded tail;
       Separate sewer: insert a pointed tail.
       & Pointed tail. 
       (Fig.~\ref{fig:teaser}F)\\
    \hline
       How much road runoff is there in the area? (1–5 map score)
       &
       & Add stick-ons to the tail. Scores 1–2 rounded, score 3 square, score 4–5 pointed. Annotate.
       & Rounded stick-ons with \textit{``2''} note.
       (Fig.~\ref{fig:teaser}F)\\
    \hline
       How much discharge is in the recent available data? (1–5 map score)
       &
       & Add stick-ons to the tail. Scores 1–2 rounded, score 3 square, score 4–5 pointed. Annotate.
       & No data \\
    \hline
       How much misconnection is in the area? \textit{(Check the pollution health map)}
       &
       & Add stick-ons to the tail. Scores 1–2 rounded, score 3 square, score 4–5 pointed. Annotate.
       & Square stick-on with ``3'' note. 
       (Fig.~\ref{fig:teaser}F)\\
    \hline
       Has any improvement been made?
       &
       & Add rounded stick-ons and annotate
       & Park regeneration, with a new lake.\\\hline
    \hline
       \textbf{Overall evaluation}
       &
       & \textbf{Work on the head} 
       & \\
    \hline
       & \textit{How do you feel about the overall quality and health of your river?}
       & Replace head with a rounded/pointed shape for positive/negative. Draw an emoji face with your feelings.
       & Rounded head. Happy emoji face. (Fig.~\ref{fig:teaser}F)\\
    \hline
    \end{tabular}
    }
    \caption{Table of the stepped prompts to explore the River Health data set (column 1), reflect on own personal connection (column 2), correspondent action to physicalize the data (column 3), and how it shows in the example (column 4 + refer to Fig.~\ref{fig:teaser}).}
    \label{tab:encodings}
    \vspace{-10pt}
\end{table*}

%Users select an element of the data set if interest to them, then chose the appropriate core and then the set of flaps in order to represent it as a 3D glyph.

\paragraph{Personalizing the Glyphs by Encoding One's Personal Relationships to the Data Set}
As they proceed to uncover the data and craft their physicalization, participants are invited to add personal specifications to the glyph using stick-on fins and label them with personal annotations. Notes can include their reasons for choosing the river, how they access its space, and the activities they engage in around it. In addition, through the choice of positive/negative stick-on flaps and the emoji face, participants are invited to express their personal evaluation of the data and their emotional response. Depending on the participants' characteristics and the workshop setup in which the toolkit is used, they can engage in deeper reflection and personalize the template (e.g., by creating their own color coding or additional shape encodings). The personalization of the template could also take advantage of additional encodings, such as texture and patterns. Further flaps or tape strips can be attached, expanding the basic structure of the glyph. Or changes can be made to adapt the instructions and steps to one's own learning style and crafting preferences.

\paragraph{Spatialization and Contribution to a Collective Map Display}
Once the glyph sculptures are assembled, we invite the participants to place them within a spatial environment. Options range from a hanging display, where position, height, and direction offer additional opportunities to encode data, to a floor mat with a printed map of London's rivers on which the finished glyphs are spatialized. The spatialization of 3D glyph physicalizations within a larger environment opens opportunities for participants to share their individual work, appreciate others' physicalization results, and reflect comparatively. In addition, contributing to a collective display can help provide opportunities to feel part of a community and create and celebrate belonging through making.

\section{Reflection}

%Design contribution
\paragraph{Initiating Exploration of the Design Space of 3D Glyph Physicalization} Although a preliminary design and prototype, our work initiates exploration of the design space of material 3D glyphs. It extends the current two-dimensional design of glyph-based visualization to consider opportunities for further encodings in the third dimension and in the materiality of physicalizations. Thus, it expands the capacity of glyphs to visualize multi-dimensional data by adding channels offered by physicality. %(e.g., the tactility and weight of materials)
Also, distributed in three dimensions, the information escapes the risk of clattering. However, the 3D glyph cannot be fully seen at a glance, unlike its graphic counterpart. Still, they can be discovered gradually and read by turning and examining the physicalization from different angles, thus allowing one to focus on different segments of the data at each step and comprehend it incrementally. \change{Lastly, despite the resemblance to objects, it is important to underline the abstract and symbolic nature of the space of 3D glyphs, especially when the exploration of material glyphs is motivated by public engagement: for example, in our case, to distinguish it from a geographic representation of the river network.}

%Public engagement contribution
\paragraph{Public Engagement Contribution} Data crafting activities, through toolkits or workshops, provide participants with opportunities to engage hands-on in exploring data with many cognitive, reflective, and societal benefits~\cite{Jansen:2015:CHI}. Recent work by Peng et al.~\cite{Peng:2026:CHI:phys4good} reviews several projects on data physicalization through the lens of social good. Our suggested use of glyph-based physicalization of data for public engagement is based on signs and symbols, tapping into a conceptual mode of representing and communicating information that is fundamental to human nature~\cite{Borgo:Glpyphs:2012}. Thus, glyph physicalizations are promising for engaging participants, transcending cultural boundaries, social barriers, and varying levels of technical expertise. %yet communicate it in immediate and approachable ways that help engage the public.
Furthermore, our work suggests that glyph-based physicalizations contribute to the public engagement conversation by enabling both in-depth data discovery and personal identification. Thus, bridging between cognition and reflection and helping the data set and its significance come to life (see the following paragraphs).
%They provide cognitive, reflective and emotional opportunities for comprhending data. 

%On one end, 3d glyphs offer an articulate form of physical representation that helps users to digest the complexity of the data.On the other end, they allow to integrate self-mapping

\paragraph{Scaffolding Cognition, Learning, and Critical Thinking over a Complex Data Set}
Engagement with data and theory in physical form is common to didactic experiences in schools or museums, mixing interaction and spectacularization with educational and display outcomes. A strength of physicalizations is their ability to communicate data persuasively. However, physicalizations are also often associated with physical constraints imposed on the data set~\cite{Jansen:2015:CHI}. By contrast, 3D glyphs offer an articulate form of physical representation, able to incorporate a rich complexity of data variables, to the advantage of users' comprehension and interpretation. By supporting complexity, it guides an in-depth exploration of the data. In our prototyping of the 3D glyph kit, we observed that the template helps users digest complex information and keep track of concepts and steps as they explore different angles of a data set. 3D glyphs help scaffold the process of exploring the data set, discovering its value and meaning, and breaking it down into small cognitive steps and the corresponding physicalizing actions. 

%Supporting congitive process, studyign an unfamiliar data-set, tracking and mapping one's learning onto the glyph as information is gradually acquired from the data set. 
%Keeping track while moving away from the map on to explore external links.
%Embedding personal viewpoints and own connections/relationship to the data set as we go. Supporting subjective learning and own reasoning.
%Potential to use this 3d glyph making as a studying and learning methodology.

\paragraph{Integrating Self-Mapping, Personal Reflection, and Emotional Connection to the Data}
Peng et al. outline among the challenges of public engagement, the public's \textit{``struggle to develop a sense of connection to the phsyicalization''}, causing them to see it \textit{``as someone else's story''}~\cite{Peng:2026:CHI:phys4good}.
Our glyph physicalization template strives to combine exploration of an external data set with users' information from their personal experience data. Data physicalization is an effective tool for self-mapping and self-reflection over personal data~\cite{thudt_data_2017}, for example, through constructive input visualizations~\cite{Huron:2014:Constructive-vis}, or bullet journaling~\cite{yu_creatively_2025}, or self-mapping through crafting~\cite{thudt_data_2017, wannamaker_data_nodate}. Our template for making 3D glyphs aims to incorporate self-mapping strategies into the exploratory process, prompting participants to reflect on the personal significance of the data set and its broader topic in relation to their experiences. This reflective result is strengthened by the opportunities to handle the glyphs by hand, label them, and write on them. The glyph physicalization template makes integrated use of annotation, building on Bae et al.'s invitation to treat annotation \textit{``as a holistic, integrated element''}, thereby aiding the interpretation of data representations. In our work, annotations facilitate a dialog between the participant and the data set, supporting reflection on one's own habits and behavior in relation to blue spaces. The further use of emojis encourages the incorporation of an emotional response to the topic of river health into the physical representation.

\section{\change{Limitations and} Prospected Work}
\change{We presented the preliminary design and prototype of a toolkit for engaging participants in making 3D glyph fish representations of River Health in London; however, at this stage the template is generalized and has not yet been tested with users in the wild.} The next steps in our work aim to deploy the design and prototype of the glyph physicalization kit in a public engagement workshop for London Data Week 2026. This event advocates for public engagement with data. On this occasion, we expect to observe the public's approaches to creating river health data glyphs. We intend to address the effects of glyph-making on learning about river health, understanding how the overarching topic of climate change affects Londoners, and assessing its impact on raising awareness, inviting environmentally responsible attitudes, and fostering a sense of belonging and community.

In future work, we also expect to see variations in the 3D glyph template, maintaining its fundamental slot-in grammar and fish-like structure, but varying materials, size, and the level of personalization, \change{thus situating it to engage different audiences and contexts. The founding template we presented can support the work of data designers and facilitators adapting it for use in the wild. For example, we envision the toolkit may be used in schools, encouraging pupils to source the crafting material from repurposed domestic waste. Or it could be distributed as a self-led activity sheet and a basic kit of materials in libraries, social venues, and community centers. Further variations to the toolkit's materials may be targeted to involving climate change skeptics, exploring more sensorially enticing materials and aesthetically pleasant outcomes of the glyph fish and of its collective display.}

%While our prototype ends at 1 round of slot-in shapes attached to the core, to each slot-in shape there is the potential to attach further shapes, growing the physicalization from the centre towards its periphery in all directions. Thus creating a structure that can be adapted to representing various types of data, even more complex anda variate. 

%Further modifications to materials can be made to engage diverse audiences and adapt the template for use in various learning and creative contexts (per number of participants, location/setup, level of cretaive skills, level of data knowledge...)

%Potential expansions: 
%e.g. attach flaps to flaps --> branch out and grow infinitely
%e.g. they can personalize the code e.g. creating more flaps of other shapes, adding color of choice (according to an own personal code of their creation)

\section{Conclusion}
\change{Overall, in this paper, we} presented a design and prototype of a 3D glyph-based data physicalization toolkit to create opportunities for public engagement with London river health data, an issue exacerbated by progressive climate change. We articulated opportunities for \change{material} 3D glyphs to scaffold the cognitive exploration of complex data sets, while allowing users to self-map and incorporate personal experience and emotion into the data set. \change{We encourage the use of the toolkit in public engagement workshops, situating it for different settings and tuning it to the characteristics of diverse groups. We} outline expectations for the template to help foster a personal connection to and a sense of belonging in blue spaces, thereby encouraging care in the wake of climate change.

%%Acknowledgments section.
\begin{acks}
This work was supported by a training grant from the UK Research and Innovation: EPSRC DTP Studentship.
\change{We would like to thank Mike Brondbjerg and Paul Hodgson from the Greater London Authority (GLA) for their valuable insights and assistance in helping us understand the London Datastore and the River Health Map}
\end{acks}

%%
%%Bibliography style and bibliography file.
\bibliographystyle{ACM-Reference-Format}
\bibliography{3D-glyphs}

@String{Computing = "Computing" }

@String{Computer = "{IEEE} Computer" }

@misc{Alliance-world-scientists-warnings,
    author = {{Alliance of World Scientists}},
    title = {{Journal Articles Related to Scientists' Warning}},
    year = "2024",
    howpublished = "https://scientistswarning.forestry.oregonstate.edu/journal-articles-related-scientists-warning",
    note = "(accessed: 06.05.2026)"
}

@inproceedings{Borgo:Glpyphs:2012,
    booktitle = {Eurographics 2013 - State of the Art Reports},
    editor = {M. Sbert and L. Szirmay-Kalos},
    title = {{Glyph-based Visualization: Foundations, Design Guidelines, Techniques and Applications}},
    author = {Borgo, Rita and Kehrer, Johannes and Chung, David H. S. and Maguire, Eamonn and Laramee, Robert S. and Hauser, Helwig and Ward, Matthew and Chen, Min},
    year = {2013},
    publisher = {The Eurographics Association},
    DOI = {10.2312/conf/EG2013/stars/039-063},
    address = {Girona, Spain},
    pages = {39--63}
}

@book{Clarke-GLA:2002:Londond-warming-report,
    author = {Clarke, Simon and Kersey, Jim and Trevorrow, Emily and Wilby, Rob and Shackley, Simon and Turnpenny, Jon and Wright, Andy and Hunt, Alistair and Crichton, David},
    title = {London's Warming: The Impacts of Climate Change on London},
    publisher = {London Climate Change Partnership},
    year = {2002},
    address = {London}
}

@misc{GLA:London-Datastore,
   author = {{Greater London Authority}},
    title = {{London Datastore}},
    year = "2026",
    howpublished = "https://data.london.gov.uk",
    note = "(accessed: 11.05.2026)"
}

@misc{GLA:river-health-website,
    author = {{Greater London Authority}},
    title = {{River Health}},
    year = "2026",
    howpublished = "https://www.london.gov.uk/programmes-strategies/environment-and-climate-change/climate-change/climate-adaptation/river-health?ac-235446=235445",
    note = "(accessed: 08.05.2026)"
}

@inproceedings{Huron:2014:Constructive-vis,
    author = {Samuel Huron and Sheelagh Carpendale and Alice Thudt and Anthony Tang and Michael Mauerer},
    title = {Constructive visualization},
    booktitle = {Proceedings of the 2014 conference on Designing interactive systems},
    year = {2014},
    pages = {433-442},
    publisher = {Association for Computing Machinery},
    address = {New York, NY, USA},
    doi = {10.1145/2598510.2598566}
}

@inproceedings{Huron:Lets-get-physical:2017,
    author = {Huron, Samuel and Gourlet, Pauline and Hinrichs, Uta and Hogan, Trevor and Jansen, Yvonne},
    title = {Let's Get Physical: Promoting Data Physicalization in Workshop Formats},
    booktitle = {Proceedings of the 2017 Conference on Designing Interactive Systems},
    year = {2017},
    pages = {1409--1422},
    publisher = {Association for Computing Machinery},
    address = {Edinburgh United Kingdom},
    doi = {10.1145/3064663.3064798}
}

@inproceedings{Jansen:2015:CHI,
    author = {Yvonne Jansen and Pierre Dragicevic and Petra Isenberg and Jason Alexander and Abhijit Karnik and Johan Kildal and Sriram Subramanian and Kasper Hornb{\ae}k},
	title = {Opportunities and Challenges for Data Physicalization},
	booktitle = {Proceedings of the 2015 CHI Conference on Human Factors in Computing Systems},
	year = {2015},
	pages = {3227–36},
	doi = {10.1145/2702123.2702180},
    publisher = {Association for Computing Machinery},
    address = {New York, NY, USA},
	note = {Seoul Republic of Korea: ACM, 2015},
}

@article{Jin:2012:hydrology-model-thames,
    author = {Jin, Li and Whitehead, Paul G. and Futter, Martyn N. and Lu, Zunli},
    title = {Modelling the impacts of climate change on flow and nitrate in the River Thames: assessing potential adaptation strategies},
    journal = {Hydrology Research},
    year = {2012},
    volume = {43},
    issue = {6},
    pages = {902-916},
    doi = {10.2166/nh.2011.080}
}

@inproceedings{Nissen:2015:data-things,
    author = {Bettina Nissen and John Bowers},
    title = {{Data-Things}: Digital Fabrication Situated within Participatory Data Translation Activities},
    booktitle = {Proceedings of the 2015 CHI Conference on Human Factors in Computing Systems},
    year = {2015},
    publisher = {ACM},
    pages = {2467-2476},
    doi = {10.1145/2702123.2702245},
    address = {New York, NY, USA}
}

@inproceedings{Peng:2026:CHI:phys4good,
    author = {Peng, Yechun and Wu, Runxi and Cao, Nan and Shi, Yang},
    title = {From Touch to Change: Understanding Public Engagement in Data Physicalization for Social Good},
    booktitle = {Proceedings of the 2026 CHI Conference on Human Factors in Computing Systems},
    year = {2026},
    pages = {1-18},
    doi = {10.1145/3772318.3790877},
    publisher = {Association for Computing Machinery},
    address = {New York, NY, USA},
	note = {Barcelona, Spain: ACM, 2026},
}

@article{Ripple:2022:warning-of-climate-emergency,
    author = {Ripple, William J and Wolf, Christopher and Gregg, Jillian W and Levin, Kelly and Rockström, Johan},
    title = {World Scientists’ Warning of a Climate Emergency 2022},
    journal = {BioScience},
    year = {2022},
    volume = {72},
    issue = {7},
    pages = {1149-1155},
    doi = {10.1093/biosci/biac083}
}

@inproceedings{van_koningsbruggen_metaphors_2024,
    author = {Van Koningsbruggen, Rosa and Haliburton, Luke and Rossmy, Beat and George, Ceenu and Hornecker, Eva and Hengeveld, Bart},
    title = {Metaphors and `{Tacit}' {Data}: the {Role} of {Metaphors} in {Data} and {Physical} {Data} {Representations}},
    booktitle = {Proceedings of the Eighteenth International Conference on Tangible, Embedded, and Embodied Interaction},
    publisher = {ACM},
    month = feb,
    year = {2024},
    pages = {1--17},
    doi = {10.1145/3623509.3633355},
    address = {New York, NY, USA}
}

@article{Wang:2019:Emotional-response,
    author = {Yun Wang and Adrien Segal and Roberta Klatzky and Daniel Keefe and Petra Isenberg and Jorn Hurtienne and Eva Hornecker and Tim Dwyer and Stephen Barrass},
    title = {An Emotional Response to the Value of Visualization},
    journal = {IEEE Computer Graphics and Applications},
    year = {2019},
    volume = {39},
    issue = {5},
    pages = {8-17},
    doi = {10.1109/MCG.2019.2923483}
}

@misc{wiki:dataphys,
   author = {Pierre Dragicevic and Yvonne Jansen},
   title = {{Data Physicalization Wiki}},
   year = "2021",
   howpublished = "\url{http://dataphys.org}",
   note = "(accessed: 21.04.2024)"
 }

@inproceedings{de_kreij_data_2024,
    address = {Delft Netherlands},
    title = {Data {Physicalization} and {Tangible} {Manipulation} for {Engaging} {Children} with {Data}: {An} {Example} with {Air} {Quality} {Data}},
    isbn = {9798400704420},
    shorttitle = {Data {Physicalization} and {Tangible} {Manipulation} for {Engaging} {Children} with {Data}},
    url = {https://dl.acm.org/doi/10.1145/3628516.3655788},
    doi = {10.1145/3628516.3655788},
    language = {en},
    urldate = {2024-08-09},
    booktitle = {Proceedings of the 23rd {Annual} {ACM} {Interaction} {Design} and {Children} {Conference}},
    publisher = {ACM},
    author = {De Kreij, Sander and Ranasinghe, Champika and Degbelo, Auriol},
    month = jun,
    year = {2024},
    pages = {507--516},
}

@inproceedings{hogan_pedagogy_2017,
    address = {Edinburgh United Kingdom},
    title = {Pedagogy \& {Physicalization}: {Designing} {Learning} {Activities} around {Physical} {Data} {Representations}},
    isbn = {978-1-4503-4991-8},
    shorttitle = {Pedagogy \& {Physicalization}},
    url = {https://dl.acm.org/doi/10.1145/3064857.3064859},
    doi = {10.1145/3064857.3064859},
    language = {en},
    urldate = {2023-11-01},
    booktitle = {Proceedings of the 2017 {ACM} {Conference} {Companion} {Publication} on {Designing} {Interactive} {Systems}},
    publisher = {ACM},
    author = {Hogan, Trevor and Hinrichs, Uta and Jansen, Yvonne and Huron, Samuel and Gourlet, Pauline and Hornecker, Eva and Nissen, Bettina},
    month = jun,
    year = {2017},
    pages = {345--347},
}

@inproceedings{thudt_data_2017,
    title = {Data craft: integrating data into daily practices and shared reflections},
    shorttitle = {Data craft},
    url = {https://research-repository.st-andrews.ac.uk/handle/10023/10910},
    language = {eng},
    urldate = {2024-08-09},
    author = {Thudt, Alice and Hinrichs, Uta and Carpendale, Sheelagh},
    month = may,
    year = {2017},
    note = {Accepted: 2017-06-06T10:30:14Z},
    booktitle = {CHI Workshop on Quantified Data and Social Relationships},
    publisher = {ACM},
    address = {Denver, Colorado, United States},
    pages = {1--5}
}

@inproceedings{aragon_risingemotions_2021,
    address = {Virtual Event Italy},
    title = {{RisingEMOTIONS}: {Bridging} {Art} and {Technology} to {Visualize} {Public}’s {Emotions} about {Climate} {Change}},
    isbn = {978-1-4503-8376-9},
    shorttitle = {{RisingEMOTIONS}},
    url = {https://dl.acm.org/doi/10.1145/3450741.3465259},
    doi = {10.1145/3450741.3465259},
    language = {en},
    urldate = {2026-05-11},
    booktitle = {Creativity and {Cognition}},
    publisher = {ACM},
    author = {Arag\'{o}n, Carolina and Jasim, Mahmood and Mahyar, Narges},
    month = jun,
    year = {2021},
    pages = {1--10},
}

@inproceedings{sauve_edo_2023,
    address = {Warsaw Poland},
    title = {Edo: {A} {Participatory} {Data} {Physicalization} on the {Climate} {Impact} of {Dietary} {Choices}},
    isbn = {978-1-4503-9977-7},
    shorttitle = {Edo},
    url = {https://dl.acm.org/doi/10.1145/3569009.3572807},
    doi = {10.1145/3569009.3572807},
    language = {en},
    urldate = {2026-05-13},
    booktitle = {Proceedings of the {Seventeenth} {International} {Conference} on {Tangible}, {Embedded}, and {Embodied} {Interaction}},
    publisher = {ACM},
    author = {Sauvé, Kim and Dragicevic, Pierre and Jansen, Yvonne},
    month = feb,
    year = {2023},
    pages = {1--13},
}

@article{perovich_chemicals_2021,
    title = {Chemicals in the {Creek}: designing a situated data physicalization of open government data with the community},
    volume = {27},
    copyright = {https://ieeexplore.ieee.org/Xplorehelp/downloads/license-information/IEEE.html},
    issn = {1077-2626, 1941-0506, 2160-9306},
    shorttitle = {Chemicals in the {Creek}},
    url = {https://ieeexplore.ieee.org/document/9233460/},
    doi = {10.1109/TVCG.2020.3030472},
    language = {en},
    number = {2},
    urldate = {2026-05-13},
    journal = {IEEE Transactions on Visualization and Computer Graphics},
    author = {Perovich, Laura J. and Wylie, Sara Ann and Bongiovanni, Roseann},
    month = feb,
    year = {2021},
    pages = {913--923},
}

@misc{web:londondatastore10,
   author = {Joseph Colombeau},
   title = {10 years of the {London Datastore} \& thinking on city data for the next decade},
   year = "2020",
   howpublished = "\url{https://chiefdigitalofficer4london.medium.com/10-years-of-the-london-datastore-thinking-on-city-data-for-the-next-decade-b634ae62dc3c}",
   note = "(accessed: 14.05.2026)"
 }

@misc{web:londondatastore2025,
   author = {Martine Wauben},
   title = {Local data sharing: what has worked well, and what challenges do we still see in London?},
   year = "2025",
   howpublished = "\url{https://data.london.gov.uk/blog/local-data-sharing-what-has-worked-well-and-what-challenges-do-we-still-see-in-london/}",
   note = "(accessed: 14.05.2026)"
 }

@inproceedings{yu_creatively_2025,
    address = {Bordeaux/Talence France},
    title = {Creatively {Supporting} {Mental} {Wellbeing}: {A} {Tangible} {Toolkit} to {Scaffold} {Self}-{Tracking} through {Mindful} {Colouring}},
    isbn = {979-8-4007-1197-8},
    shorttitle = {Creatively {Supporting} {Mental} {Wellbeing}},
    url = {https://dl.acm.org/doi/10.1145/3689050.3704944},
    doi = {10.1145/3689050.3704944},
    language = {en},
    urldate = {2025-05-09},
    booktitle = {Proceedings of the {Nineteenth} {International} {Conference} on {Tangible}, {Embedded}, and {Embodied} {Interaction}},
    publisher = {ACM},
    author = {Yu, Jingxin and Ayobi, Amid and Marshall, Paul and O'Kane, Aisling Ann},
    month = mar,
    year = {2025},
    pages = {1--17},
}

@misc{wannamaker_data_nodate,
    title = {Data {Embroidery}: {Exploring} {Alternative} {Mediums} for {Personal} {Physicalization}},
    language = {en},
    author = {Wannamaker, Kendra and Oehlberg, Lora and Carpendale, Sheelagh and Willett, Wesley},
    note = {(conference poster)},
    year = {2019}
}

@inproceedings{sauve_econundrum_2020,
    address = {Eindhoven Netherlands},
    title = {Econundrum: {Visualizing} the {Climate} {Impact} of {Dietary} {Choice} through a {Shared} {Data} {Sculpture}},
    isbn = {978-1-4503-6974-9},
    shorttitle = {Econundrum},
    url = {https://dl.acm.org/doi/10.1145/3357236.3395509},
    doi = {10.1145/3357236.3395509},
    language = {en},
    urldate = {2026-05-14},
    booktitle = {Proceedings of the 2020 {ACM} {Designing} {Interactive} {Systems} {Conference}},
    publisher = {ACM},
    author = {Sauvé, Kim and Bakker, Saskia and Houben, Steven},
    month = jul,
    year = {2020},
    pages = {1287--1300},
}

@article{ellwein_using_2014,
    title = {Using {Rich} {Context} and {Data} {Exploration} to {Improve} {Engagement} with {Climate} {Data} and {Data} {Literacy}: {Bringing} a {Field} {Station} into the {College} {Classroom}},
    volume = {62},
    issn = {1089-9995, 2158-1428},
    shorttitle = {Using {Rich} {Context} and {Data} {Exploration} to {Improve} {Engagement} with {Climate} {Data} and {Data} {Literacy}},
    url = {https://www.tandfonline.com/doi/full/10.5408/13-034},
    doi = {10.5408/13-034},
    language = {en},
    number = {4},
    urldate = {2026-05-14},
    journal = {Journal of Geoscience Education},
    author = {Ellwein, Amy L. and Hartley, Laurel M. and Donovan, Sam and Billick, Ian},
    month = nov,
    year = {2014},
    pages = {578--586},
}

@article{karyda_narrative_2021,
    title = {Narrative {Physicalization}: {Supporting} {Interactive} {Engagement} {With} {Personal} {Data}},
    volume = {41},
    issn = {1558-1756},
    shorttitle = {Narrative {Physicalization}},
    url = {https://ieeexplore.ieee.org/abstract/document/9200790/authors#authors},
    doi = {10.1109/MCG.2020.3025078},
    number = {1},
    urldate = {2023-12-08},
    journal = {IEEE Computer Graphics and Applications},
    author = {Karyda, Maria and Wilde, Danielle and Kjærsgaard, Mette Gislev},
    month = jan,
    year = {2021},
    note = {Conference Name: IEEE Computer Graphics and Applications},
    pages = {74--86},
}

\end{document}